\documentclass[twocolumn,showpacs,preprintnumbers,amsmath,amssymb,
superscriptaddress]{revtex4}

\usepackage[dvips]{color}

\usepackage{graphicx}
\usepackage{dcolumn}
\usepackage{bm}
\usepackage{dsfont}

\begin{document}

\preprint{APS/123-QED}
\title{Magnetically tuned spin dynamics resonance}

\author{J. Kronj{\"a}ger}
\affiliation{Institut f{\"u}r Laserphysik, Universit{\"a}t
Hamburg, Luruper Chaussee 149, D-22761 Hamburg, Germany.}
\author{C. Becker}
\affiliation{Institut f{\"u}r Laserphysik, Universit{\"a}t
Hamburg, Luruper Chaussee 149, D-22761 Hamburg, Germany.}
\author{P. Navez}%
\affiliation{Labo Vaste-Stoffysica en Magnetisme, Katholieke
Universiteit Leuven, Celestijnenlaan 200D, B-3001 Leuven, Belgium.}%
\affiliation{additional address: Universit{\"a}t Duisburg-Essen,
Universit{\"a}tsstrasse 5, 45117 Essen, Germany}
\author{K. Bongs}
\affiliation{Institut f{\"u}r Laserphysik, Universit{\"a}t
Hamburg, Luruper Chaussee 149, D-22761 Hamburg, Germany.}
\author{K. Sengstock}
\affiliation{Institut f{\"u}r Laserphysik, Universit{\"a}t Hamburg,
  Luruper Chaussee 149, D-22761 Hamburg, Germany.}

\date{\today}
\begin{abstract}
  We present the experimental observation of a magnetically tuned
  resonance phenomenon resulting from the spin mixing dynamics of ultracold atomic gases.
  In particular we study the magnetic field dependence of spin
  conversion in F=2 $^{87}$Rb spinor condensates in the crossover
  from interaction dominated to quadratic Zeeman dominated
  dynamics. We discuss the observed phenomenon in the framework
  of spin dynamics as well as matter wave four wave mixing. Furthermore
  we show that the validity range of the single mode approximation for
  spin dynamics is significantly extended in the regime of high magnetic field.
\end{abstract}

\maketitle

Ultracold quantum gas spin mixtures are currently receiving
rapidly growing experimental and theoretical interest. They
combine the unprecedented control achieved in single component
Bose-Einstein condensates with intrinsic degrees of freedom. The
spin-dependent coupling connects quantum gas physics to
fundamental magnetic phenomena. Earlier work analyzed the
basic magnetic properties of these spinor
condensates~\cite{Ho1998a,Ohmi1998a,Law1998a,Koashi2000a} and
demonstrated antiferromagnetic behavior in the $F=1$ state of
$^{23}$Na~\cite{Stenger1999a} and the $F=2$ state of
$^{87}$Rb~\cite{Schmaljohann2004a} as well as ferromagnetic
behavior in the $F=1$ state of
$^{87}$Rb~\cite{Schmaljohann2004a,Chang2004a}.

A particularly interesting feature of these clean and undisturbed
systems is that they give access to quantum aspects of magnetism such
as interaction-driven spin
oscillations~\cite{Schmaljohann2004a,Chang2004a,Kuwamoto2004a,Schmaljohann2004b,Higbie2005a,Widera2005a,Chang2005a},
the quantum classical transition~\cite{Kronjaeger2005a} or domain
formation~\cite{Stenger1999a,Chang2005a,Sadler2006a}. The magnetic field dependence
of the underlying spin conversion process has been analyzed in detail
for $F=1$ spinor
condensates~\cite{Pu2000a,Romano2004a,Zhang2005a,Kronjaeger2005a} and
besides oscillatory behavior a resonance phenomenon was
predicted~\cite{Zhang2005a}. Experiments so far focused on magnetic
fields above the resonance ~\cite{Kronjaeger2005a} and on magnetic
phase controlling of the spinor dynamics~\cite{Chang2005a} but a spin
mixing resonance has not yet been observed.

\begin{figure}[t]
\resizebox*{\columnwidth}{!}{\includegraphics{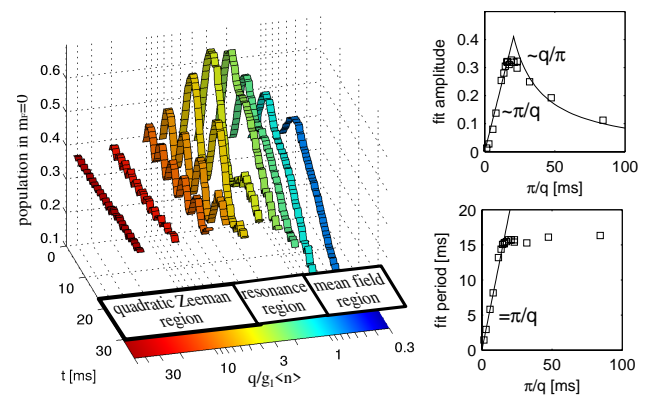}}
\caption{\label{fig:resonance} Left: Temporal evolution of the
$m_F=0$-population for different ratios of quadratic Zeeman and
interaction energy (see text). Right: Magnetic field dependence of
the amplitude and period of the $m_F=0$ population as extracted
from a sinusoidal fit to the oscillatory part of the data.}
\end{figure}

In this letter we study for the first time magnetic field tuned spin
mixing in $F=2$ $^{87}$Rb spinor condensates. In particular as a main
result of this letter we demonstrate resonant spin mixing behavior in
spinor condensates. This resonance phenomenon becomes evident in the
temporal population evolution shown in Fig.~\ref{fig:resonance}. As a
surprising central effect, the mean field driven spin dynamics
(population transfer from $m_F\pm 1 $ to $m_F=0$) strongly depends on
the external magnetic field. In contrast to studies on the natural
ground state phase of the system and early findings on spin dynamics
suppression~\cite{Schmaljohann2004b,Kuwamoto2004a} at high magnetic
field values we now find, that an external magnetic field not only
leads to adverse effects but can even stimulate spin dynamics. The
spin mixing amplitude is not large for zero magnetic offset field, but
shows a pronounced maximum for finite magnetic field values as
predicted in~\cite{Zhang2005a,Kronjaeger2005a}. The observation and
understanding of this resonance in self driven spin dynamics in
quantum gases is central for detailed magnetic state manipulation.  A
quantum control for general spin systems (for F=1
see~\cite{Chang2005a}) based on this spinor resonance process might
lead to new schemes for spin entanglement and quantum information
applications with spinor gases.

The experimental setup has been described in~\cite{Schmaljohann2004a}
and the procedure resembles the one in~\cite{Kronjaeger2005a}. We
prepare degenerate $^{87}$Rb ensembles in the $F=2$, $m_F=-2$ state
containing typically $3\cdot 10^5$ atoms in an optical dipole trap
(trapping frequencies $\omega_x : \omega_y :\omega_z \approx 2\pi\cdot
14,5:105:600\,s^{-1}$).
A radio frequency $\pi/2$-pulse of typically 40$\mathrm{\mu s}$
duration rotates this state to our initial state
$\zeta^{\text{ini}}(0) = (1/4, 1/2, \sqrt{6}/4, 1/2,
1/4)^T$~\footnote{We use the notation $\psi= \sqrt{ n}\left(
    \zeta_{+2}, \zeta_{+1}, \zeta_0, \zeta_{-1}, \zeta_{-2}(t)
  \right)^\top $ for the vector order parameter of the spinor
  condensate.}.
After holding the sample for a variable time, $t$, at a well
defined magnetic offset field we switch off the trapping
potential. The population in each spin component is extracted from
absorption images after Stern-Gerlach separation during time of
flight~\cite{Stenger1999a,Schmaljohann2004a}.
Fig.~\ref{fig:resonance} shows the temporal evolution of the
$m_F=0$-population in dependence of the relative value of the
quadratic Zeeman frequency, $q$ and the mean field spin mixing
frequency, $g_1\langle n \rangle $. The magnetic field dependence
is parameterized to first order by $q = p^2/\omega_\text{hfs}$
with $p= \mu_B B/2 \hbar$ and the $^{87}$Rb ground state hyperfine
splitting $\omega_{\text{hfs}} \approx 2\pi\cdot 6.835\,
\mathrm{GHz}$. The spin dependent mean field interaction is
proportional to the density, $n$, of the sample and for $^{87}$Rb
dominated by the parameter $g_1$ introduced below.

Our observations show that the whole spin mixing dynamics depends
critically on the relative size of the energies associated to the
quadratic Zeeman shift and the spin dependent mean field
interaction. Interestingly the oscillation amplitude in
Fig.~\ref{fig:resonance} is reduced whenever one of these energies
dominates, i.e. in both the quadratic Zeeman regime
($q>>g_1\langle n \rangle $) and the mean field regime
($q<<g_1\langle n \rangle $). A maximum occurs close to balanced
quadratic Zeeman and mean field contributions. The important point
is, that these two effects compete with each other. The quadratic
Zeeman effect lifts the energy of the $m_F=0$ state with respect
to the other states, while the mean field interaction lowers it.
More precisely the phase evolution due to these energy
contributions is occurring in opposite directions. However,as spin
mixing is a coherent process, this phase determines the direction
of spin mixing. The accumulation of many particles in one state
can thus only happen for slow phase evolution, i.e. close to the
balanced case.

We want to emphasize, that the many-body spin mixing resonance
presented in this letter crucially relies on nonlinear
many-particle effects like self- and cross-phase modulation. It is
therefore fundamentally different from the "Rabi"-resonance case
observed in spin-changing two-particle
collisions~\cite{Widera2005a}, in which these effects are absent.

In the following we will begin our analysis of the observed spin
mixing resonance phenomenon with a discussion of the the limiting
cases. In particular we present and interpret analytic solutions
for the asymptotic behavior in these regimes. Our analysis is
based on the evolution due to the mean field hamiltonian developed
in~\cite{Koashi2000a} using the single mode approximation (SMA).

Spin mixing occurs as a consequence of two-particle collisions and
can be classified by the total spin $f$ of the colliding pair. For
$F=2$ the allowed values are $f=0,2,4$. The spin mixing
interactions are parameterized by the coupling coefficients
$g_1=\frac{4\pi \hbar^2}{m}\frac{a_4-a_2}{7}$ and $g_2=\frac{4\pi
\hbar^2}{m}\frac{4}{5}\frac{7a_0-10a_2+3a_4}{7}$~\cite{Koashi2000a},
using the spin-dependent s-wave scattering lengths $a_f$. The
dynamics of $F=2$ spinor condensates will in general be a complex
superposition of the different mixing processes summarized in
table~\ref{tab:mixing} (we use the notation $(i)$ for
waves/particles in the $m_F=i$ state). This table also lists the
relative quadratic Zeeman energy shifts of the involved states.
\begin{table}
\caption{\label{tab:mixing} Four wave mixing processes in $F=2$ spinor BEC.}
\begin{ruledtabular}
\begin{tabular}{lcr}
process & coupling & q.Z. energy diff.\\
\hline $(0) + (0) \leftrightarrow (+1) +(-1)$ & $g_1$ & 2q\\
$ (+1) + (+1) \leftrightarrow (+2) + (0)$ & $g_1$ & 4q\\
$(-1) + (-1) \leftrightarrow (0) + (-2)$ & $g_1$ & 4q\\
$(+1) + (-1) \leftrightarrow (+2) + (-2)$ & $g_1$ & 6q\\
\hline $(0) + (0)\leftrightarrow (+1) + (-1)$ & $g_2$ & 2q\\
$(+1) +(-1) \leftrightarrow (+2) + (-2)$ & $g_2$ & 6q\\
$(0) + (0) \leftrightarrow (+2) + (-2)$ & $g_2$ & 8q\\
\end{tabular}
\end{ruledtabular}
\end{table}
$^{87}$Rb has the advantage that the value of $g_2$ is very
small~\cite{Widera2006a}, such that the processes listed in the
lower part of table~\ref{tab:mixing} is largely suppressed and can
be neglected in the following~\footnote{We experimentally checked
the validity of this assumption by confirming that the initial
state $\bm{\zeta^{(0)}}(0) = (\sqrt{6}/4, 0, -1/2, 0,
\sqrt{6}/4)^T$ shows no visible spin mixing within the first
15\,ms, which is the relevant timescale for coherent evolution in
our experiment~\cite{Kronjaeger2005a}. For this initial state all
$g_1$-couplings vanish and the remaining coupling $(0)+(0)
\leftrightarrow (+2) + (-2)$ only depends on
$g_2$~\cite{Saito2005a}.}. In addition we will concentrate on the
specific initial state $\zeta^{\text{ini}}(0)$ used in our
experiments. For the population evolution in the limit of low
magnetic field ($q<<g_1\langle n \rangle $), i.e. the ''mean field
region'', we find to first order:
\begin{eqnarray}\label{eq:mf}
|\zeta_0^{mf}|^2 &=& \frac{3}{8}\left( 1- \frac{q}{2 g_1 \langle n
\rangle } \left[ cos(4g_1\langle n \rangle t)-1 \right] \right) \nonumber \\
|\zeta_{\pm 1}^{mf}|^2 &=& \frac{1}{4}\left( 1+ \frac{q}{2 g_1
\langle n \rangle } \left[ cos(4g_1\langle n \rangle t)-1 \right] \right) \nonumber \\
|\zeta_{\pm 2}^{mf}|^2 &=& \frac{1}{16}\left( 1- \frac{q}{2 g_1
\langle n \rangle } \left[ cos(4g_1\langle n \rangle t)-1 \right]
\right)
\end{eqnarray}
This mean field dominated evolution is just a sinusoidal
oscillation between the spin components and shows a magnetic field
independent period $T_{mf}=\frac{\pi }{2g_1 \langle n \rangle }$
as observed in Fig.~\ref{fig:resonance}. Furthermore these
equations indicate that the oscillation amplitude in this regime
should be proportional to the quadratic Zeeman shift $q$, which is
reflected in Fig.~\ref{fig:resonance}.

In the other limit at high magnetic field, i.e. the quadratic
Zeeman regime with $q>>g_1\langle n \rangle $, the population
evolution is dominated by the quadratic Zeeman energy and we find:
\begin{eqnarray}\label{eq:qz}
|\zeta_0^{qZ}|^2 &=& \frac{3}{8}\left\{ 1- \frac{ g_1\langle n
\rangle}{2q}\left[ 2 \left( \cos(2qt)-1 \right) +
\frac{\cos(4qt)-1}{2} \right] \right\} \nonumber \\
|\zeta_{\pm 1}^{qZ}|^2 &=& \frac{1}{4}\left\{ 1+ \frac{ g_1\langle
n \rangle}{2q}\left[ \frac{3\left( \cos(2qt)-1 \right) }{4}  -
\frac{\cos(6qt)-1}{12} \right] \right\} \nonumber \\
|\zeta_{\pm 2}^{qZ}|^2 &=& \frac{1}{16}\left\{ 1+ \frac{
g_1\langle n \rangle}{2q}\left[ \frac{3\left( \cos(4qt)-1 \right)
}{2}  \right. \right. \nonumber \\
& & \left. \left. + 3 \left( \cos(2qt)-1 \right) +
\frac{\cos(6qt)-1}{3} \right] \right\} .
\end{eqnarray}
The oscillation amplitude in this regime decays with $\frac{1}{q}$
in accordance to the data in Fig.~\ref{fig:resonance}. However now
several frequencies are involved as a result of the different
mixing processes in $F=2$ spin dynamics. The oscillation periods
reflect the different quadratic Zeeman shifts for the possible
$g_1$ mixing processes as listed in table~\ref{tab:mixing}. For
the chosen initial state, spin mixing is dominated by the
$\cos(2qt)$ term  in Eq.~\ref{eq:qz} connected to the $(0) + (0)
\leftrightarrow (-1) + (+1)$ process, which has a factor of four
higher weight than the $\cos(4qt)$ term.

From the above tendencies in the population oscillation, it is
clear, that a maximum amplitude must occur at an intermediate
magnetic field value, i.e. the resonance found in
Fig.~\ref{fig:resonance}.

\begin{figure}[t]
\resizebox*{\columnwidth}{!}{\includegraphics{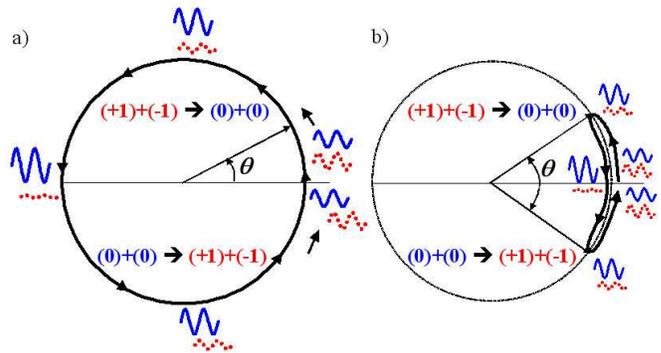}}
\caption{\label{fig:fourwaveanalogy} Schematic representation of
four wave mixing in a) the quadratic Zeeman regime  and b) the
mean field regime. The direction of spin mixing is indicated in
the upper and lower half of the circle symbolizing the relative
phase $\theta $. For $\theta
>0 $ the $(0)$ waves (upper waves) get populated, while for
$\theta <0$ the $(+1)$ and $(-1)$ wave (lower waves) increase.}
\end{figure}

In the following we will develop some more qualitative physical
insights by an interpretation of the resonance phenomenon in terms
of ''phase matching'' in four wave mixing
(see Fig.~\ref{fig:fourwaveanalogy}). First we note that given the
prefactor $\frac{1}{16}$ the oscillation amplitude of the $m_F=\pm
2$ wave is relatively small in both the mean field regime and the
quadratic Zeeman regime. This justifies an approximate view as a
pure $(-1) + (+1) \leftrightarrow (0) + (0)$ four wave mixing
process. In contrast to nonlinear optics or four wave mixing in
single component condensates~\cite{Deng1999a} we do not deal with
wavevector or momentum modes~\cite{Burke2004a,Goldstein1999a} but with spin
modes, i.e. here the $m_F=+1$
wave and the $m_F=-1$ wave couple to two times the $m_F=0$ wave
(this view is justified, as we are considering trapped samples in
a single momentum state). Spin mixing in trapped samples
corresponds to degenerate collinear four wave
mixing~\cite{Goldstein1999a} and (in contrast to single component
four wave mixing) in principle allows infinite interaction times,
only limited by finite temperature
effects~\cite{Lewandowski2003a,Erhard2004a,Schmaljohann2004b,Mur2006a,Kronjaeger2005a}.

In view of the optical analogy it is obvious, that phase matching
considerations are essential to understand the resonance in spinor
four wave mixing. The important point is, that the value of the
relative phase of the initial and final waves/spin components
determines the direction of wave mixing. In our case the $(0)$
wave will be populated if the combined phase of the $(+1) + (-1)$
waves is ahead of twice the $(0)$ wave phase, i.e. $\theta =
\varphi_{+1}+ \varphi_{-1}-2\varphi_{0} \in [0..\pi]$ and it will
be depopulated, if $\theta \in [\pi .. 2\pi]$ (modulo $2\pi $).

In general the relative phase evolution of the spinor components
in spin mixing is highly nontrivial, as in addition to the
quadratic Zeeman energy shifts it depends on the spin coupling and
the spin component populations in a nonlinear way. We find that
for our system, the competition between mean field driven
dephasing (tending to decrease $\theta $) and quadratic Zeeman
shift driven dephasing (tending to increase $\theta $) determines
the evolution of the system. Most importantly the evolution
depends on the relative size of the quadratic Zeeman energy shift
and the maximally achievable mean field shifts, as shown in
Fig.~\ref{fig:fourwaveanalogy}. For large magnetic fields, i.e.
always negligible mean field energy shifts, $\theta $ will
continuously grow, i.e. the $(\pm 1)$ wave evolves faster than the
$(0)$ wave, and the population transfer shows an oscillatory
behavior depending on which wave is lagging behind at which
instant (Fig.~\ref{fig:fourwaveanalogy}a). In this regime the
oscillation period is expected from Eq.~\ref{eq:mf} to be given by
$\frac{\pi}{q}$. Furthermore the oscillation amplitude, which
depends on the unidirectional mixing time, should be proportional
to this period. Both these expectations are confirmed by the data,
as shown in the inset of Fig.~\ref{fig:resonance}.

For small magnetic fields the mean field energy shift grows
with increasing population in the $(0)$ wave, until it exceeds
the quadratic Zeeman shift and thus reverses the evolution
of $\theta $, i.e. the $(0)$ wave will speed up and decrease its
lag behind the $(\pm 1)$ wave, eventually overtaking it. In this
case $\theta $ will remain confined in the interval $[-\pi ..
\pi]$ and oscillate around $\theta =0$ (
Fig.~\ref{fig:fourwaveanalogy}b). In this regime the population
oscillation amplitude is expected to decrease with decreasing $q$,
i.e. at lower magnetic field, as less population transfer is
necessary to create a mean field energy of the size of the
quadratic Zeeman shift. This decrease in amplitude is also
confirmed by the data in Fig.~\ref{fig:resonance}.

In the resonance region at intermediate magnetic field, the
relative movement of the $(0)$ and $(\pm 1)$ waves is very slow,
with $\theta $ being on the border of oscillation and continuous
increase. This leads to long time unidirectional spin mixing and
thus maximum amplitude.

We want to point out that also the initial direction of spin
mixing increasing the $(0)$ wave amplitude as observed in
Fig.~\ref{fig:resonance} can be explained by simple arguments. The
chosen initial state $\zeta^{\text{ini} }$ has zero phase shift
$\theta =0$ and shows no mean field dephasing between the $(\pm
1)$ and $(0)$ waves. This is due to the fact, that it originates
from the fully stretched $(-2)$ eigenstate subjected to a 90
degree rotation). As the mean field part of the Hamiltonian is
invariant under rotations $\zeta^{\text{ini}}$ stays a mean field
eigenstate. Consequently the quadratic Zeeman energy shift will
always cause the $(\pm 1)$ wave to evolve faster then the $(0)$
wave or in other words $\theta $ will initially always grow. The
transfer of particles going in the direction of the wave lagging
behind will thus result in an initial increase of the $(0)$ wave
(see Fig.~\ref{fig:fourwaveanalogy}).

In addition to the understanding of the tendencies observed in
Fig.~\ref{fig:resonance} within the single mode approximation, a
quantitative comparison allows to deduce the validity limits of
this approximation for the elongated trap geometry used in the
experiment. The breakdown of the single mode approximation is
accompanied by the strong damping of the oscillation at low $q$,
which is probably due to the onset of spatial structure formation in spin
dynamics~\cite{Higbie2005a}. In contrast to the quadratic Zeeman
dominated regime spin dynamics in the mean field regime is
characterized by a density dependent coupling constant $g_1
\langle n \rangle $. For mean field driven spin dynamics the
higher density parts of the spinor condensate will thus dephase
relative to the lower density parts and population oscillations in
the total fractions will be washed out. 

\begin{figure}[t]
\includegraphics{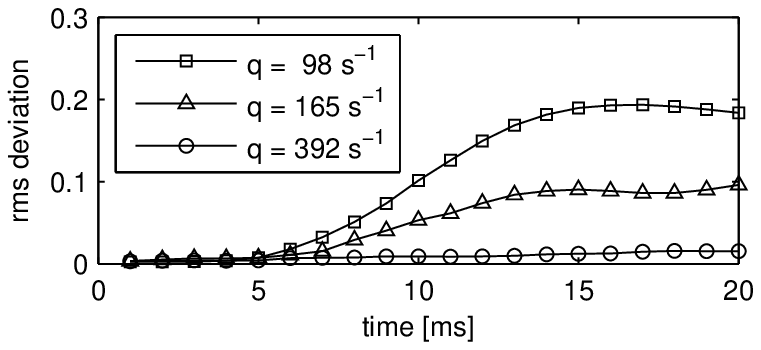}\\
\includegraphics{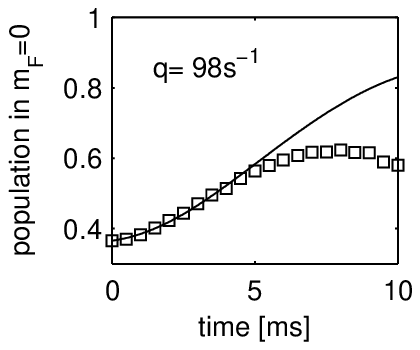}\includegraphics{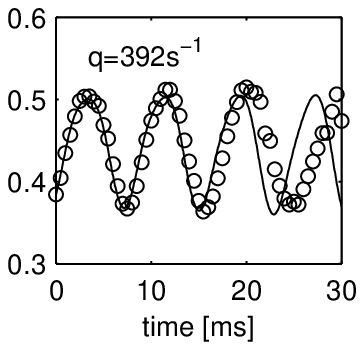}
\caption{\label{fig:smacompare} TOP: Deviation (rms of $m_F=0$ population) of SMA theory fit ($g_1\langle n \rangle = 54\,\mathrm{s^{-1}}$ fixed) from our data (Fig.~\ref{fig:resonance}). The abscissa specifies the time range over which the fit was perfomed. BOTTOM: two examples of fits compared to data.}
\end{figure}

The validity range of the SMA can also be inferred from our
experimental data (Fig.~\ref{fig:smacompare}). While the evolution of
Spinor condensates in our elongated geometry starts to deviate from
SMA behavior after $\approx 5\,$ms in the mean field regime, there is
a near perfect agreement over more than 20\,ms in the mean field
region.  Our most important observation in this respect is that even
for condensate sizes much above the spin healing length the single
mode approximation works remarkably well once the quadratic Zeeman
effect dominates the relative phase evolution.

In conclusion we have investigated a magnetically tunable
resonance in coherent spin dynamics and have analyzed this
phenomenon in terms of phase matching considerations, which are
applicable to general values of total spin. In the future this
resonance will allow the controlled use of spin dynamics to fully
convert one spinor state into another, even with local control by
shifting in and out of the validity range of the single mode
approximation. These possibilities open new perspectives for
spatiotemporal tailoring of the spinor condensate wavefunction,
e.g. for the creation of highly nonclassical spin states with
spatially varying degree of entanglement.

This work was funded in part by Deutsche Forschungsgemeinschaft
(DFG) within SPP 1116. P.N. acknowledges support from the German
AvH foundation and from the Junior Fellowship program of the
KULeuven Research Council.


\end{document}